# Registry-Dependent Potential for interfaces of water with graphene

Zhicheng Feng,[1] Yuanpeng Yao,[1] Jianxin Liu,[1] Bozhao Wu[1], Ze Liu,[1,2] Wengen Ouyang,[1,2*]

[1]*Department of Engineering Mechanics, School of Civil Engineering, Wuhan University, Wuhan, Hubei 430072, China.*

[2]*State Key Laboratory of Water Resources & Hydropower Engineering Science, Wuhan University, Wuhan, Hubei 430072, China.*

[*]Corresponding author. Email: w.g.ouyang@whu.edu.cn

ABSTRACT

An anisotropic interlayer potential that can accurately describe the van der Waals interaction of the water-graphene interface is presented. The force field is benchmarked against the many-body dispersion-corrected density functional theory. The parameterization of ILP yields good agreement with the reference dataset of binding energy curves and sliding potential energy surfaces for various configurations of a water molecule deposited on monolayer graphene, indicating the developed force field enhancing significantly the accuracy in the empirical description of water-graphene interfacial interactions. The water contact angles of monolayer and multilayer graphene extracted from molecular dynamics simulations based on this force field are close to the experimental measurements and predict the hydrophilic nature of graphene. The theoretical approach proposed in this work can be easily extended to mimic the van der Waals interactions between water molecules and various two-dimensional layered materials for studying their wetting properties.

## 1. Introduction

The wettability of graphene has attract great attention due to its emerging applications in various fields, such as water desalination,[1-5] water purification,[6,7] self-cleaning,[8] energy storage,[9] anti-corrosion[8] and anti-icing[10]. However, there is a long debt on the intrinsic wettability of graphene (hydrophobic or hydrophilic) in past decades and an accurate understanding remains elusive,[11,12] because the experimental measurements of the water contact angles (WCAs) for graphene are spread in a wide range of 10 °-180 °.[13-15] The reasons are mainly attributed to the surface contamination[16,17] and the effect of underneath supporting substrate.[11,12,18] In recent experiments, much effort has been put to eliminate the effect of surface contamination and supporting substrate by measuring the WCA of free-standing graphene layers with high quality in a controlled environment.[19-21] These measurements show that the clean free-standing graphene exhibiting a hydrophilic nature, with the WCA in the range of 30°-85°.[19-21]

On the other hand, theoretical predictions of WCA of graphene also spreads in a wide range (0°-129.9°)[21-30], which is resulted from the strong dependence of the WCA on the choice of interlayer potential between water and graphene. For instance, the two-body Lennard-Jones (LJ) potential ($V(r) = 4\epsilon[(\sigma/r)^{12} - (\sigma/r)^6]$) has been widely employed in MD simulations.[31-33] The parameters $\varepsilon$ and $\sigma$, determining the binding energy and equilibrium distance, are parametrized to describe the van der Waals interaction between graphene and water.[24,33-36] The typical parameters for describing the interaction of the oxygen atoms in water molecules and carbon atoms in graphene, $\varepsilon_{CO}$ and $\sigma_{CO}$, are in the range of 2.07-7.14 meV and 3.13-4.06 Å, respectively.[23,24,27,30,34] Such diverse parameters results in the scattered WCA values, which indicates the LJ potential is not a good choice for studying the wetting properties of graphene. To describe the van der Waals (vdW) interaction between water molecule and graphene more accurately, first-principles calculations such as vdW-augmented density functional theory (DFT) are more appropriate, but should be very careful for the choice of DFT method.[21,37,38] Recent studies show that the many-body electronic structure methods, such as diffusion Monte Carlo (DMC) method, coupled cluster theory (CCSD(T)) and the random phase approximation (RPA) predict water-graphene interaction strength with sub-chemical accuracy.[39] However, these DFT methods are extremely time-consuming and cannot applied directly for calculating the WCA of graphene. Thus, an accurate and more efficient theoretical description of the intrinsic wetting property of large-scale graphene remains a challenge.

To overcome this barrier, a possible way is to develop a more accurate force field in the framework of MD simulations. Inspired by the successful application of registry-dependent interlayer potential (ILP) to the two-dimensional (2D) layered materials,[40-46] we modified the original ILP for the description of water-graphene interaction in this work. We first performed dispersion corrected density functional theory (DFT) calculations for the graphene−water heterojunction, using the Perdew-Burke-Ernzerhof (PBE) exchange−correlation functional within the generalized gradient approximation, that augmented by a nonlocal many-body dispersion (MBD-NL) treatment of long-range correlation. This DFT method is much more efficient than that of DMC method, but with acceptable accuracy (see Table 1 and Sec. 3 of the SI). According to our previous experience on developing the interlayer force fields for 2D materials, both the binding energy (BE) curves and sliding potential surfaces (PESs) for the graphene-water heterojunction should be benchmarked to get reliable parameters. The sliding PES has never been emphasized in previous theoretical studies for water-carbon interfaces, however, this is very important for simulating accurately the transport/tribological properties of water droplets on graphitic surfaces. The ILP is carefully parameterized against to the DFT reference data and shows good agreement with all of the DFT data sets. It should be noted that we failed to parameterize the LJ potential to the same DFT reference data, which further indicates that the LJ potential cannot describe accurately the vdW interaction of water-graphitic interfaces. As a benchmark test of the developed ILP, we performed MD simulations using this force field for large-scale water-graphene heterostructures, and calculated the water contact angles (WCAs) of monolayer and multilayer graphene as 54° and 44°, respectively, which falls well within the experimental range.[16,19-21,47] The ILP developed in this work thus provides a more unified and accurate description of the water-graphene interaction, which can be easily extended to various 2D materials. Our approach opens the avenue for studying the wetting properties of various 2D materials more accurately.

## 2. Methodology

*2.1. DFT Method*

The reference DFT data are obtained using the MBD-NL augmented PBE functional[48], as implemented in the FHI-AIMS code,[49] with the tier-2 basis-set[50] using tight convergence settings including all grid divisions and a denser outer grid. Relativistic effects are neglected. The supercell size of the water/graphene models is set as $6a \times 6a \times 1$ ($a$ = 2.463 Å is the lattice constant of graphene), for all configurations, a vacuum size of 100 Å along $z$-direction was used with a $k$-points mesh of 9×9×1. Convergence of the DFT results with respect to the supercell size, the vacuum size and the $k$-points was established (see Sec. S1 in the Supporting Information (SI)).

## 2.2. Model Systems

The model system consists of a water molecule depositing on periodic graphene surface. Initially, we chose six configurations of water/graphene models, in which the water molecule is deposited on the graphene surface with different stacking modes and orientations, as shown in Sec. S2 of the SI. Then we optimized these configurations with PBE+MBD-NL and got two stable configurations: i) a "two-leg" configuration (Figure 1a-b), in which the oxygen atom is atop the center of a hexagonal carbon ring with two hydrogen atoms pointing to the graphene plane and ii) a "one-leg" configuration (Figure 1c-d), in which the oxygen atom is approximately above a carbon atom with one hydrogen atom pointing to the graphene plane. Similar results have been reported in literatures.[21,38,51,52] After optimizations, the carbon-carbon bond length within the graphene sheet and the oxygen-hydrogen bond length are 1.422 Å and 0.969 Å, respectively, while the bond angle of water molecule is 104.18°. The "two-leg" and "one-leg" configurations were used to calculate the BE curves and sliding PESs hereinafter.

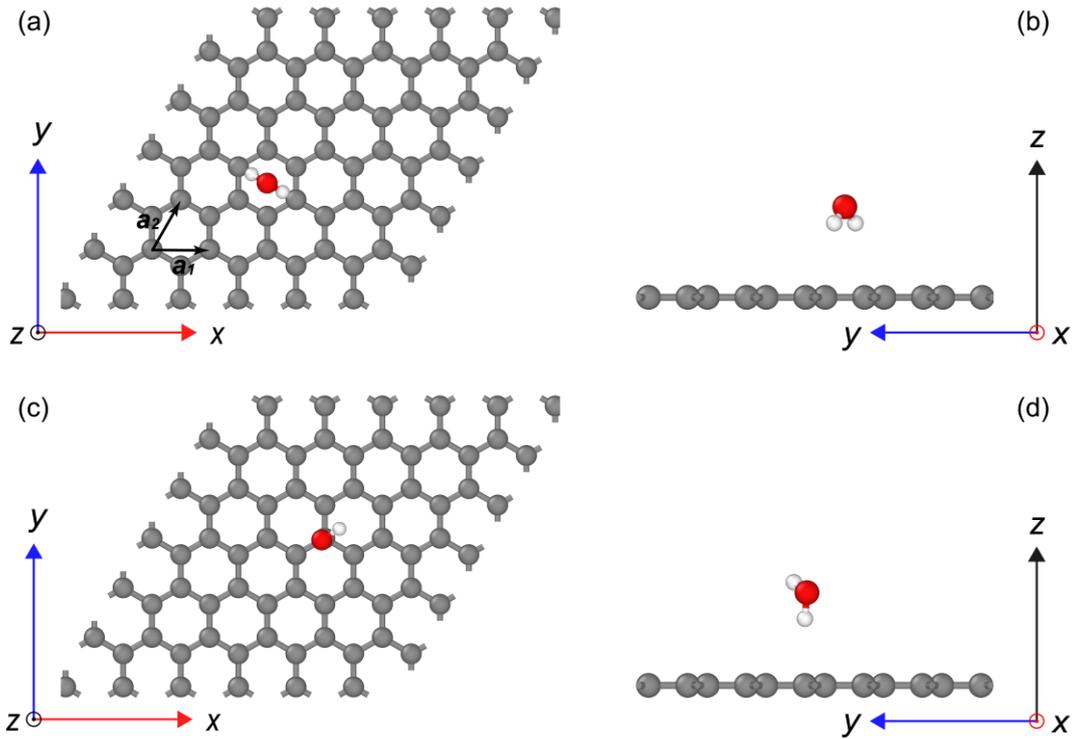

*Figure 1. Two stable configurations of water molecule atop graphene, obtained from PBE +MBD-NL optimizations. Top view (a, c) and side view (b, d) of the "two-leg" and the "one-leg" configuration, respectively. $a_1, a_2$ are the lattice vectors of the unit cell in graphene with a magnitude of 2.463 Å. The grey, red and white spheres represent carbon, oxygen and hydrogen atoms, respectively.*

## 2.3. DFT reference data

### 2.3.1. Binding Energy Curves

Figure 2a presents the BE curves of the optimized "one-leg" and "two-leg" configurations that calculated by PBE+MBD-NL, with the interlayer distance between the oxygen atom of the water molecule and the graphene layer varying from 2 Å to 16 Å. The equilibrium distances for the "two-leg" and "one-leg" configurations are 3.29 Å and 3.36 Å, respectively, which is consistent with computations for water/graphene heterostructure using various methods.[21,51,52] The corresponding binding energies of water molecule on graphene are −117.8 meV and -114.0 meV, respectively, when compared with the diffusion Monte Carlo (DMC) method, the BE values calculated by PBE+MBD-NL are overestimated by ~20 % as shown in Table 1.[39] Considering the relatively large uncertainty in the DMC calculations, the actual overestimation would be smaller.[53]

**Table 1.** *Binding energy ($E_b$) and equilibrium distances (the vertical distance between oxygen atom and graphene) for the "two-leg" and "one-leg" configurations (see Figure 1) that calculated with different levels of theory.*

|  | Two leg | | One leg | |
| --- | --- | --- | --- | --- |
| Approach | $E_b$ (meV) | Height (Å) | $E_b$ (meV) | Height (Å) |
| CCSD(T)[54] | -123 | 2.61 | -- | -- |
| MP2[54] | -106 | 2.66 | -- | -- |
| DMC[39] | -99±6 | 3.37 | -92±6 | 3.46 |
| p-CCSD(T)[39] | -87 | -- | -76 | -- |
| RPA[39] | -82±1 | 3.41 | -74±1 | 3.52 |
| RPA+GWSE[39] | -98±1 | 3.33 | -87±1 | 3.45 |
| PBE+D3[21] | -128.3 | 3.30 | -125.5 | 3.38 |
| BLYP+D3[21] | -122.0 | 3.32 | -121.4 | 3.42 |
| PBE+MBD (This work) | -117.8 | 3.29 | -114.4 | 3.36 |

In addition to the regular calculations of BE curves between water molecule and graphene, we also calculate the BE by rotating the water molecule respect to the graphene layer for three typical configurations: "one-leg" (360° rotation, Figure 2b), "two-leg" (60° rotation, Figure 2c) and parallel ("zero-leg") configuration (120° rotation, Figure 2c). The last configuration is generated by setting the water molecule be parallel to the graphene layer. During the rotation, the interlayer distance between the oxygen atom and graphene layer is fixed at 3.4 Å for the "one-leg" configuration and 3.3 Å for "two-leg" and parallel configuration, respectively. Figure 2b-c show clearly that the BE of water molecule on graphene depends strongly on its orientation, indicating the LJ potential is unable to describe the vdW interaction between water and graphene,[39] see details in Sec. 6 of the Supplementary. In contrast, the ILP fitting shows good qualitative and quantitative agreement with the DFT reference data, as is shown in Figure 2 (the solid lines).

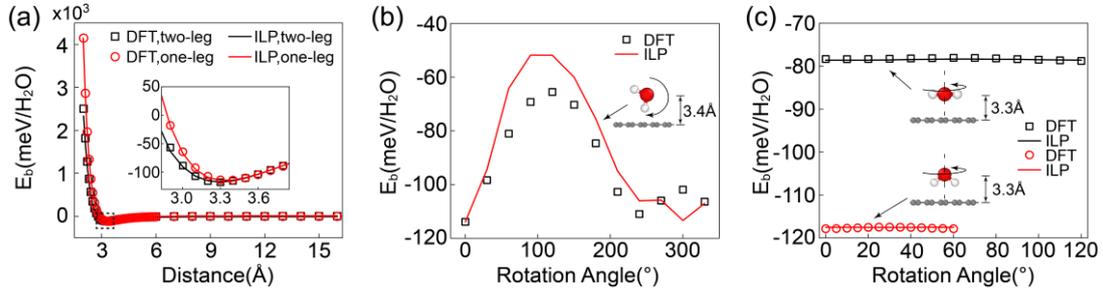

***Figure 2.*** *The BE calculations of the water molecule on monolayer graphene. (a) The BE curves of the "two-leg" configuration (black squares and lines) and "one-leg" configuration (red circles and lines), calculated using PBE+MBD-NL (open symbols), along with the corresponding ILP fitting results (solid lines). The inset in panel (a) provides zoom-in on the equilibrium interlayer distance region. (b,c) The BE as a function of the rotation angle by rotating (b) the "two-leg" configuration and (c) the "one-leg" and parallel configurations at their equilibrium interlayer distances. The inserts in panels (b,c) show the rotational axis and directions (the arrows) of the water molecules.*

*2.3.2. Sliding Potential Energy Surfaces*

Figure 3 presents the calculated sliding PESs of both "two-leg" (the top row) and "one-leg" configurations (bottom row) at their equilibrium interlayer distances by rigidly moving the water molecule atop the graphene layer using PBE+MBD-NL (Figure 3a,d) and ILP (Figure 3b,e). The corresponding differences between the reference DFT data and the ILP results are presented in Figure 3c,f. As is shown in this figure, the ILP fitting is in good qualitative and quantitative agreement with the DFT reference data. The maximal deviation between the DFT reference and ILP results of the overall sliding PESs corrugation is within 1.8 meV/$H_2O$. It should be noted that the LJ potential cannot even capture the pattern of the sliding PES (see Sec. 6 of the SI). The above results show the capability of ILP for describing the strongly anisotropic vdW interaction of the water-graphene heterostructure.

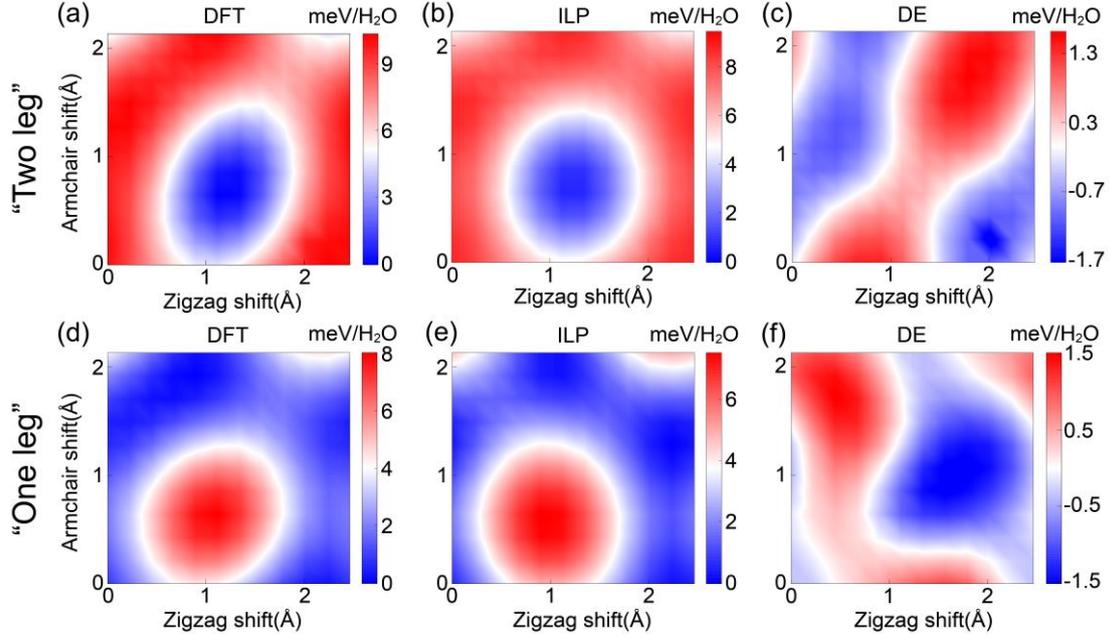

*Figure 3.* *The sliding potential energy surface (PES) of a water molecule on graphene surface, calculated at the interlayer distances of 3.3 Å and 3.4 Å for the "two-leg" (upper rows) and "one-leg" (lower rows) configurations, respectively. The left, middle, and right columns present the PES calculated using PBE+MBD-NL, ILP, and their difference, respectively.*

## 3. Force-Field Parameterization

*3.1. The formula of the interlayer potential*

In this work, the developed ILP is based on the concept of anisotropic interlayer potentials for 2D materials,[46,55-58] which consists of the following two terms: an isotropic term that describes the long-range attractive dispersive interactions and an anisotropic term that describes the Pauli-type repulsion between the graphene π electrons and the electrons in water molecule. The dispersive term treats long-range van der Waals interactions via a $C_6/r^6$ LJ type potential, dampened in the short range with a Fermi−Dirac-type function, similar to that introduced in dispersion-corrected DFT calculations to avoid double counting of correlation effects[59]:

$$E_{\text{att}}(r_{ij}) = \text{Tap}(r_{ij})\left\{-\frac{1}{1+e^{-d\left[\left(r_{ij}/(s_R \cdot r_{ij}^{\text{eff}})\right)-1\right]}} \cdot \frac{C_{6,ij}}{r_{ij}^6}\right\} \quad (1)$$

where $r_{ij}$ is the distance between a carbon atom $i$ and a hydrogen or an oxygen atom $j$. $d$ and $s_R$ are unitless parameters determining the steepness and onset of the short-range Fermi-type dampening function. $r_{ij}^{\text{eff}}$ and $C_{6,ij}$ are the sum of effective atomic radii and the pair-wise dispersion coefficients, respectively. The $\text{Tap}(r_{ij})$ function provides a continuous (up to 3rd derivative) long-range cutoff at $r_{ij} = R_{\text{cut}}$ to the potential aiming to reduce computational cost[60]:

$$\text{Tap}(r_{ij}) = \frac{20}{R_{\text{cut}}^7}r_{ij}^7 - \frac{70}{R_{\text{cut}}^6}r_{ij}^6 + \frac{84}{R_{\text{cut}}^5}r_{ij}^5 - \frac{35}{R_{\text{cut}}^4}r_{ij}^4 + 1 \tag{2}$$

According to the Kolmogorov−Crespi[58] scheme, the anisotropic (repulsive) term of the potential is constructed from a Morse-like exponential isotropic term, multiplied by an anisotropic correction with the following form:

$$E_{\text{rep}}(\mathbf{r}_{ij}) = \text{Tap}(r_{ij})e^{\alpha_{ij}\left(1-\frac{r_{ij}}{\beta_{ij}}\right)}\left[\epsilon_{ij} + C_{ij}\left(e^{-\left(\frac{\rho_{ij}}{\gamma_{ij}}\right)^2} + e^{-\left(\frac{\rho_{ji}}{\gamma_{ji}}\right)^2}\right)\right] \tag{3}$$

in which $\text{Tap}(r_{ij})$ is the cutoff smoothing function describing by eq (2). $\alpha_{ij}$ and $\beta_{ij}$ determine the slope and range of the repulsive potential, respectively, while $\gamma_{ji}$ sets the width of the Gaussian decay factors in the anisotropic correction term and thus determines the sensitivity to the transverse distance, $\rho_{ij}$, between carbon atom $i$ and hydrogen or oxygen atom $j$. $C$ and $\epsilon_{ij}$ are constant scaling factors bearing units of energy. The normalized normal vectors $\mathbf{n}_i$ (i.e., $||\mathbf{n}_i||= 1$) serve to calculate the transverse distance $\rho_{ij}$ between pairs of carbon atom $i$ and hydrogen or oxygen atom $j$.

$$\rho_{ij}^2 = r_{ij}^2 - (\mathbf{n}_i \cdot \mathbf{r}_{ij})^2$$
$$\rho_{ji}^2 = r_{ij}^2 - (\mathbf{n}_j \cdot \mathbf{r}_{ij})^2 \tag{4}$$

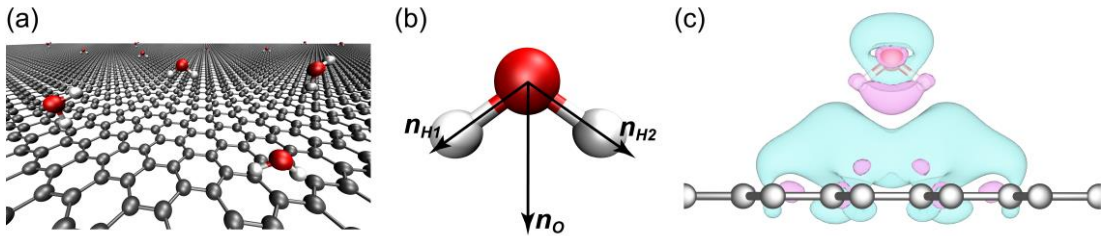

***Figure 4.*** *(a) Schematic of water molecules atop monolayer graphene, (b) The definition of normal vectors for a water molecule in the formula of ILP, (c) charge density difference of a water molecule when depositing on a graphene surface ("two-leg" configuration). Blue and purple clouds represent the accumulation and depletion of electrons, respectively.*

The atomic normal vector $\mathbf{n}_i$ defines the local normal direction to the graphene sheet (or to the water molecule) at the position of atom $i$ (Figure 4a). Here we calculate the normal vector of a carbon atom $i$ by averaging the three normalized cross products of the vectors connecting atom $i$ to its three nearest neighbors.[43,58] Figure 4c shows the charge density difference between graphene and water molecule, to account for the anisotropic nature of the water molecule, the atomic normal vectors of hydrogen atoms are assumed to lie along the corresponding oxygen-hydrogen bonds and the normal vector of oxygen atom is defined as their average (see Figure 4b), which can be expressed as follows:

$$\begin{cases} \mathbf{n}_{H_j} = \mathbf{r}_{\overrightarrow{OH_j}}/|\mathbf{r}_{\overrightarrow{OH_j}}|, & j = 1,2 \\ \mathbf{n}_O = (\mathbf{r}_{\overrightarrow{OH_1}} + \mathbf{r}_{\overrightarrow{OH_2}})/|\mathbf{r}_{\overrightarrow{OH_1}} + \mathbf{r}_{\overrightarrow{OH_2}}| \end{cases} \quad (5)$$

*3.2. Fitting Protocol*

The parameters of the ILP were fitted against reference DFT datasets ($M = M_b + M_s$) including $M_b$ BE curves and $M_s$ sliding PES as demonstrated in Sec. 2.3. Optimal ILP parameters were obtained by minimizing the following objective function that quantifies the difference between the DFT reference data and the potential predictions:

$$\Phi(\xi) = \sum_{m=1}^{M_b} w_m^b \|E_m^b(r_m, \xi) - E_m^{b,DFT}\|_2 + \sum_{m=1}^{M_s} w_m^s \|E_m^s(r_m, \xi) - E_m^{s,DFT}\|_2 \quad (6)$$

Here, $\|\cdot\|_2$ is the Euclidean norm (2-norm) that measures the difference between the ILP predictions and the DFT reference data. $\xi$ represents the set of ILP parameters. $E_m^b(r_m, \xi)$ and $E_m^s(r_m, \xi)$ represent the $M_b$ BE curves and $M_s$ sliding PES data sets, respectively. $w_m^b$ and $w_m^s$ are the corresponding weighting coefficients. The reference DFT interfacial energies, $E_m^{b,DFT}$ and $E_m^s(r_m, \xi)$, are obtained as follows: for any given configuration $m$ of the heterostructure, the total energy is first obtained from PBE+MBD-NL calculations: $E_m^{DFT,total}$. Then, the energies of the isolated graphene and water molecule, $E_m^{DFT,graphene}$ and $E_m^{DFT,water}$, are calculated using the same cell and setup as that of the composite system, respectively. The DFT interfacial energy in eq (6) is then defined as

$$E_m^{b/s,DFT} = E_m^{DFT,total} - E_m^{DFT,graphene} - E_m^{DFT,water} \quad (7)$$

The optimization was carried out using MATLAB with an interior-point algorithm[61],[62] (further details are provided in Refs. [43] and [46]). Fitted parameters and the related weights are given in Sec. S3 in the SI.

# 4. Intrinsic wettability of graphitic systems

As a typical benchmark test of the developed force field, we investigate the intrinsic wettability of multilayer graphene systems. Wettability of material is commonly characterized using the WCA. To obtain WCA of monolayer and multilayer graphene, we first implemented the ILP for water/graphene interfaces into the open-source code LAMMPS[63] and then performed MD simulations based on it. The results are visualized using the Open Visualization Tool (OVITO).[64] The water molecules are modeled using the rigid TIP4P/2005 water model,[65] which can reproduce major water properties more accurately than other commonly used models especially when considering the surface tension of water. Here the TIP4P/2005 water model consists of a Lennard-Jones (LJ) center on the oxygen (O) with $\varepsilon_{OO} = 0.7749 \text{ kJ} \cdot \text{mol}^{-1}$ (8.031 meV) and $\sigma_{OO} = 3.1589$ Å and three fixed point charges on the fourth massless M site ($-1.1128\ e$) and the two hydrogen atoms ($0.5564\ e$), the cutoff is set as 8.5 Å. The SHAKE algorithm[66] is employed to keep the rigidity of the structure. The long-range charge-charge electrostatic interactions between water molecules are calculated using the partical-partical partical-mesh (PPPM) algorithm[67] with an accuracy of $10^{-5}$ and a real space cutoff of 10 Å.

The graphene is modeled via the second generation of REBO potential[68] and the interaction between water molecules and graphene is simulated using the developed ILP. In the MD simulations, the droplets (1536-9464 water molecules) are put initially in the center of the graphene layers, which are periodic in *x* and *y* directions. The size of the supercell is 21.329 nm×19.703 nm (each graphene layer contains 16000 carbon atoms), which is large enough to avoid self-interactions of water molecules for the largest droplets in our simulations. For all the MD simulations, a time step of 2.0 fs is used, which is small enough to get convergent results.[35]

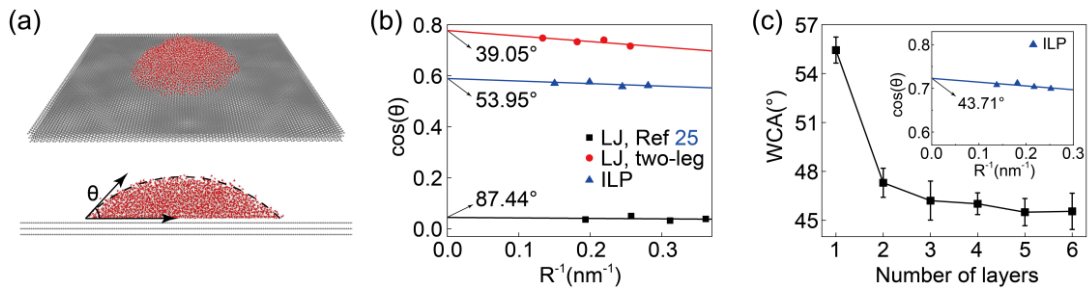

*Figure 5.* MD simulation results based on the developed ILP. (a) The spatial distribution of 4000 water molecules on three layers of graphene. (b) The calculated WCA (θ) for droplets with various sizes on monolayer graphene, from which the contact angle $\theta_{\text{inf}}$ for an infinitely large droplet can *be obtained using eq (8). The slopes of solid lines are related to the line tension of water on graphene (see eq (8)). (c) The influence of the number of graphene layers on the WCA (herein 4000 water molecules were used). The error bars were estimated as the standard deviation of the measured results in the last 5 ns. The inset in panel (c) provides the size dependence of the contact angle for a droplet depositing on six layers of graphene.*

## 4.1. The WCA for Monolayer Graphene

We first simulate the droplet/graphene system to calculate its WCA. It's known that the WCA of graphene depends on the size of depositing droplet.[69] In order to determine the macroscopic contact angle $\theta_{\text{inf}}$, the model systems with different numbers of water molecules ($N_w$) on the surface of monolayer graphene were considered: 1536, 2312, 4000, 9464. The droplets were first relaxed under the NVT ensemble for 6 ns with utilizing the Nosé-Hoover thermostat[70] to control the temperature at 300 K, while the graphene substrate is kept rigid, we checked that the flexibility of the graphene layer has a negligible effect on WCA calculated using ILP (see the discussion in Sec. 5 of the SI). To avoid the error caused by the shape distortion of water droplets during the dynamic simulations, we began to collect the data when the water droplet reaches equilibrium. Herein the density profiles of the water droplet at the last 1 ns were collected. According to the radial density profile of the water droplet, we can fit a spherical cap shape which yields the WCA (see the details for calculating WCA in Sec. S4 of the SI.). After calculations for systems containing different droplet sizes, the size-dependent WCAs to the line-tension modified Young equation can be fitted[71-74]:

$$\cos(\theta_R) = \cos(\theta_{\text{inf}}) - \frac{1}{\gamma_{\text{lv}}}\frac{\tau}{R} \qquad (8)$$

where $R$ is the average radius of the contact area between droplet and graphene, $\tau$ is the line tension, $\gamma_{\text{lv}}$ is the surface tension between liquid and vapor ($\tau$ and $\gamma_{\text{lv}}$ don't need to be known for a fit), $\theta_R$ is the measured WCA for a given droplet and $\theta_{\text{inf}}$ is the WCA of the infinite droplet. By fitting eq (8), we obtained the $\theta_{\text{inf}} = 53.95°$ for an infinite large droplet on graphene based on our ILP, confirming the hydrophilicity of graphene.[20,21] As a comparison, we also parameterized the LJ potential against the DFT reference data to describe the van der Waals interactions between water molecules and graphene. As a result, the LJ potential cannot fit the BE curves of the "two-leg" and "one-leg" configurations simultaneously, not to mention fitting the sliding PESs. Thus we parameterized the LJ potential against the BE curves of "two-leg" and "one-leg" configuration separately (see Sec. 6 of the SI) and get two sets of parameters (Table S2). The sliding PESs predicted by both sets of parameters deviate significantly from the DFT reference data (see Figure S5 in SI).

With these LJ potential parameters, the same MD simulations as the above were carried out. The $\theta_{\text{inf}}$ is calculated as 39.05° (Figure 5b) and 0° when the LJ parameters are parameterized against to the BE curve of the "two-leg" and "one-leg" configuration, respectively. The latter case indicates that the droplets spread to a flat plane on the surface of graphene. Due to the apparent discrepancy of $\theta_{\text{inf}}$ using two sets of LJ parameters, it is further verified that the LJ description of the Van der Waals interactions between water and graphene is not reliable.

*4.2. The WCA for Multilayer Graphene*

We further investigate the effect of substrate thickness on the WCA by increasing the number of graphene layers from one to six. The graphene layers are AB-stacked with an interlayer distance of 3.4 Å. The number of water molecules on graphene layers in the simulations is chosen as 4000. Similar to above simulations, the droplets were equilibrated for 10 ns with the NVT ensemble at 300 K, while the graphene layers are kept at rest (see more details in Sec. 4.1). The density profile data is collected in the last 5 ns, in which the WCA is calculated during each nanosecond. Figure 5c shows the averaged WCA gradually decrease with increasing the number of graphene layers, and finally converges to ~45° when the number of graphene layers reaches five. To obtain WCA for an infinite large droplet on six layers of graphene, we used the same treatment for monolayer graphene as shown in Sec. 4.1. After fitting line-tension modified Young equation (eq (8)), we extracted $\theta_{\text{inf}} = 43.71°$ for an infinitely large droplet on six layers of graphene as shown in the inset of Figure 5c.

*Table 2. A summary of the reported Water contact angle (WCA) values of graphene in experiments and simulations, which includes the following information: WCA and its corresponding sample used in experiments (Samples), method used to determine the WCA in experiments (Method). Setup (Setup) and force field (Force field / Force field parameters) used to describe interaction between graphene and water in simulations. Parameters $\sigma_{CO}$ (Å), $\varepsilon_{CO}$ (meV), $\sigma_{CH}$ (Å), $\varepsilon_{CH}$ (meV) of truncated 12–6 LJ potential are given. Parameters of other potentials are omitted.*

| | WCA | Samples | Method | Refs |
|---|---|---|---|---|
| Experiments | 143.2° | 1L[a] graphene (GO reduction)/glass | Sessile drop (use an optical CA meter) | [75] |
| | 91°[b] | 1L graphene/Cu | Sessile drop | [76] |
| | 33.2° | 1L graphene/Si | Sessile drop (axisymmetric drop-shape analysis profile (ADSA-P) method) | [11] |
| | 78.8° | 1L graphene/Au | | |
| | 86.2° | 1L graphene/Cu | | |
| | 48.1° | 1L graphene/glass | | |
| | 72.9°±1.27° | 1L graphene/SiC | circle-fit method | [77] |
| | 88° | 1L graphene/SiO$_2$ | Sessile drop (use CMOS camera and ESEM) | [78] |

| | | WCA | Setup | Force field / Force field parameters | |
|---|---|---|---|---|---|
| | | 68±1°[b] | 1L graphene/SiO$_2$ | advancing WCA( use Kruss DSA100 goniometer) | [19] |
| | | 85°±1°[b] | partially suspended graphene(1L) | | |
| | | 62.4°±0.9° | HOPG | | |
| | | 51°±5°[b] | 1L graphene/sapphire | captive bubble method | [47] |
| | | 180°[b] | free-standing graphene (FG) | "liquid marble" experiment | [14] |
| | | 42°±3°[b] | FG | captive bubble method | [20] |
| | | ~45°[b] | 4L FG | | |
| | | 91±1°[b] | 1L graphene/polydimethylsiloxane | Sessile drop | [13] |
| | | 10°±2°[b] | 1L graphene/agarose hydrogel | | |
| | | 30°±5°[b] | 1L graphene/ice | | |
| | | 30°[b] | FG | ESEM characterization | [21] |
| | | 45°±3° | HOPG | Sessile drop | [79] |
| | | 35° | HOPG | ESEM characterization | [80] |
| | | 64.4° | HOPG | Sessile drop | [16] |
| | | 69° | HOPG | Sessile drop | [81] |
| | | 65° | HOPG | Sessile drop | [17] |
| | | 75.2-83.2° | graphite particles | Sessile drop | [82] |
| MD Simulations | LJ potential | **WCA** | **Setup** | **Force field / Force field parameters** | - |
| | | 0° | TIP3P water model/graphite(0001) | $\sigma_{CO}$= 3.19, $\varepsilon_{CO}$= 4.05, $\sigma_{CH}$= 2.82, $\varepsilon_{CH}$= 2.62 | [22] |
| | | 82.6°±0.9° | SPC/E water model/FG(2L) | $\sigma_{CO}$= 3.19, $\varepsilon_{CO}$= 4.6 | [32] |
| | | 129.9° | SPC/Fw water model/FG(1L) | $\sigma_{CO}$= 3.19, $\varepsilon_{CO}$= 2.07 | [23] |
| | | 111.1° | SPC/Fw water model/FG(1L) | $\sigma_{CO}$= 3.58, $\varepsilon_{CO}$= 2.07 | [83] |
| | | 36.4°±3.5° | SPC/E water model/FG(2L) | $\sigma_{CO}$= 3.44, $\varepsilon_{CO}$= 3.69, $\sigma_{CH}$= 2.69, $\varepsilon_{CH}$= 1.66 | [24] |
| | | 89°±0.5° | TIP4P/2005 water model/FG(1L) | $\sigma_{CO}$= 4.01, $\varepsilon_{CO}$= 3.00 | [35] |
| | | 94.9° | SPC/E water model/FG(1L) | $\sigma_{CO}$= 3.19, $\varepsilon_{CO}$= 4.06 | [30] |
| | | 91.9° | SPC/E water model/FG(2L) | | |
| | | 98.9° | SPC/E water model/FG(4L) | | |
| | | 90.4° | SPC/E water model/FG(8L) | | |
| | | 45.7°±1.3° | TIP3P water model/FG(1L) | $\sigma_{CO}$= 3.35, $\varepsilon_{CO}$= 4.48, $\sigma_{CH}$= 1.98, $\varepsilon_{CH}$= 2.46 | [84] |
| | | 19.1°±2.5° | TIP3P water model/FG(4L) | | |
| | | 88.3° | SPC/E water model/FG(1L) | $\sigma_{CO}$= 3.19, $\varepsilon_{CO}$= 4.06 | [29] |
| | | <10° | SPC/E water model/FG(2L) | $\sigma_{CO}$= 3.13, $\varepsilon_{CO}$= 7.14, $\sigma_{CH}$= 2.45, $\varepsilon_{CH}$= 1.07 | [24] |
| | | 26° | SPC water model/FG(1L) | $\sigma_{CO}$= 3.43, $\varepsilon_{CO}$= 6.07 | [21] |
| | | 100.7°± 0.4° | SPC/E water model/FG(1~6Lfrom | $\sigma_{CO}$= 3.19, $\varepsilon_{CO}$= 3.70 | [31] |
| | | 91.3°± 0.3° | | | |

| | | 89.7°± 0.3° | up to down) | | |
|---|---|---|---|---|---|
| | | 89.6°± 0.2° | | | |
| | | 90.4°± 0.3° | | | |
| | | 90.2°± 0.3° | | | |
| | | 89.6° | SPC/E water model/FG(6L) | $\sigma_{CO}$= 4.06, $\varepsilon_{CO}$= 4.43 | [27] |
| | Other potential | 56° | coarse-grained water/ FG(1L) | Morse potential | [26] |
| | | 86° | BLYPSP-4F water/ FG(1L) | Buckingham potential | [25] |
| | QMD | 87° | freestanding graphene(1L) | - | [85] |
| | | 74° | freestanding graphene(1L) | - | [28] |
| | This work | 53.95° | TIP4P/2005 water/FG(1L) | see Table S1 in SI | - |
| | | 43.71° | TIP4P/2005 water/ FG(6L) | see Table S1 in SI | - |

[a]L refers to the number of layers; [b]The airborne contamination is carefully minimized or removed in these measurements

Table 2 summarizes the reported WCAs measured (calculated) in experiments (simulations) of monolayer and multilayer graphene for different samples (setups) with various methods (force fields). As can be seen in Table 2, the experimental values of WCAs for graphene and graphite are in ranges of 10°-180° and 35°-83.2°, respectively. The WCAs measured in experiments are still very scattered even after the airborne contamination has been minimized or removed.[13,14,19-21,47,76] Reasons for the inconsistency might be attributed to the quality of the samples and WCA measuring methods.[86]

Similarity, the WCAs obtained in MD simulations scattered in the range of 0°- 129.9°, which strongly depend on the choice of force fields and their parameters. For instance, the WCAs predicted by the LJ potential range from 0°- 129.9°, with a diverse set of parameters ($\varepsilon_{CO}$= 2.07-7.14 meV, $\sigma_{CO}$= 3.13-4.06 Å). We further show that the LJ potential can only describe one BE curve for a fixed configuration (see details in Sec. 6 of the SI). All these information indicates that the simple isotropic force field cannot mimic the vdW interaction between water molecule and graphene accurately. In contrast, the developed anisotropic force field in this work is able to capture all DFT reference data well with a single set of parameters, and predicts the WCAs of ~54° and ~44° for monolayer and multilayer graphene, respectively, which are close to the values of WCA for clean graphene that reported in recent experiments.[19-21,47]

## 5. Conclusion

The results presented above indicate that the proposed registry-dependent interlayer potential is able to capture well the energetics (BE curves and sliding PESs based on many-body dispersion-corrected DFT reference data) of water-graphene heterostructure with a single set of parameters, which cannot be described by the LJ potential. The calculated water contact angle of graphitic systems using the ILP in MD simulations indicate the hydrophilicity of graphene and graphite, agreeing well with the recent experiments. The successful construction of ILP for water-graphene interface opens the way for the efficient and accurate simulations of large-scale heterostructures between water and graphitic systems, such as their structural, wetting and transport properties. The method presented in this work can be extended easily to investigate the wettability of various 2D layered materials.

## Acknowledgments

This work was supported by the Natural Science Foundation of Hubei Province (2021CFB138) and the National Natural Science Foundation of China (Nos. 12102307, 11890673 and 11890674).